%% file: main.tex
\documentclass[conference]{IEEEtran}
\IEEEoverridecommandlockouts
\input{content/def}

\def\BibTeX{{\rm B\kern-.05em{\sc i\kern-.025em b}\kern-.08em
		T\kern-.1667em\lower.7ex\hbox{E}\kern-.125emX}}

\fancypagestyle{cfooter}{ %
	\fancyhf{} 
	\cfoot{\small{© 2022 IEEE. Personal use of this material is permitted. Permission from IEEE must be obtained for all other uses, in any current or future media, including reprinting/republishing this material for advertising or promotional purposes, creating new collective works, for resale or redistribution to servers or lists, or reuse of any copyrighted component of this work in other works.}
	}
	
}

\begin{document}
	
	\title{Variational Autoencoder Leveraged MMSE\\Channel Estimation
		\thanks{This work is funded by the Bavarian Ministry of Economic Affairs, Regional Development and Energy within the project 6G Future Lab Bavaria. The authors acknowledge the financial support by the Federal Ministry of Education and Research of Germany in the program of “Souverän. Digital. Vernetzt.”. Joint project 6G-life, project identification number: 16KISK002}
	}
	
	\author{\IEEEauthorblockN{Michael Baur, Benedikt Fesl, Michael Koller, and Wolfgang Utschick}
		\IEEEauthorblockA{School of Computation, Information and Technology, Technical University of Munich, Germany}
		Email: \{mi.baur, benedikt.fesl, michael.koller, utschick\}@tum.de
	}

	\maketitle
	\thispagestyle{cfooter}
	\begin{abstract}
		We propose to utilize a variational autoencoder (VAE) for data-driven channel estimation. The underlying true and unknown channel distribution is modeled by the VAE as a conditional Gaussian distribution in a novel way, parameterized by the respective first and second order conditional moments. As a result, it can be observed that the linear minimum mean square error (LMMSE) estimator in its variant conditioned on the latent sample of the VAE approximates an optimal MSE estimator. Furthermore, we argue how a VAE-based channel estimator can approximate the MMSE channel estimator. We propose three variants of VAE estimators that differ in the data used during training and estimation. First, we show that given perfectly known channel state information at the input of the VAE during estimation, which is impractical, we obtain an estimator that can serve as a benchmark result for an estimation scenario. We then propose practically feasible approaches, where perfectly known channel state information is only necessary in the training phase or is not needed at all. Simulation results on 3GPP and QuaDRiGa channel data attest a small performance loss of the practical approaches and the superiority of our VAE approaches in comparison to other related channel estimation methods.
	\end{abstract}

	\input{content/introduction}
	\input{content/system}
	\input{content/model}
	\input{content/results}
	\input{content/conclusion}
	\vfill
	\bibliographystyle{IEEEtran}
	\bibliography{main}
	
\end{document}

%% file: content/def.tex
\usepackage[utf8]{inputenc}
\usepackage{tikz}
\usepackage{pgfplots}
\usepackage{bm}
\usepackage{amsmath} 
\usepackage{amsfonts}
\usepackage{graphicx}
\usepackage{bm}
\usepackage{siunitx}
\usepackage{cite}
\usepackage{textcomp}
\usepackage{xcolor}
\usepackage{fancyhdr}
\usepackage{algorithmic}

\pgfplotsset{
	discard if not/.style 2 args={
		x filter/.code={
			\edef\tempa{\thisrow{#1}}
			\edef\tempb{#2}
			\ifx\tempa\tempb
			\else
			\fi
		}
	}
}
\pgfplotsset{compat=1.15}

\newcommand{\vdel}{\bm{\delta}}
\newcommand{\vmu}{\bm{\mu}}
\newcommand{\vsig}{\bm{\sigma}}
\newcommand{\veps}{\bm{\varepsilon}}
\newcommand{\va}{\bm{a}}
\newcommand{\vc}{\bm{c}}
\newcommand{\vg}{\bm{g}}
\newcommand{\vh}{\bm{h}}
\newcommand{\vm}{\bm{m}}
\newcommand{\vn}{\bm{n}}
\newcommand{\vx}{\bm{x}}
\newcommand{\vy}{\bm{y}}
\newcommand{\vz}{\bm{z}}

\newcommand{\mc}{\bm{C}}
\newcommand{\mf}{\bm{F}}
\newcommand{\msig}{\bm{\Sigma}}
\newcommand{\RR}{\mathbb{R}}
\newcommand{\CC}{\mathbb{C}}
\newcommand{\diff}{\mathrm{d}}
\newcommand{\jim}{\mathrm{j}}

\DeclareMathOperator{\E}{E}
\DeclareMathOperator{\diag}{diag}
\DeclareMathOperator{\KL}{D_{KL}}

\DeclareMathOperator*{\herm}{^{\mkern-1.5mu{H}}}

%% file: content/introduction.tex
\section{Introduction}
\label{sec:intro}

Machine learning (ML), and deep learning (DL) in specific, are promising candidates for further improvements of massive MIMO systems~\cite{Bjornson2019}. DL incorporates the characteristics of a communications scenario by training neural networks based on data stemming from that scenario. The acquired knowledge about the scenario can often be leveraged to outperform classical methods in typical tasks such as channel estimation. The great performance of DL-based channel estimation has been demonstrated in a large number of publications, e.g., see~\cite{Ye2018,Soltani2019,He2018a,Neumann2018}. Roughly speaking, such approaches can either be specified as end-to-end or model-based learning~\cite{He2018} or as a mixture of both. While end-to-end approaches model the system as a black box and use neural networks to learn the unknown behavior, model-based approaches rely on analytical relationships which are parameterized using neural networks.

In this work, we want to take a model-based approach for channel estimation with the help of DL. An approach closely related to ours is the recently proposed Gaussian mixture model (GMM)-based channel estimator~\cite{Koller2021icassp,Koller2022}. The estimator is proven to converge to the true conditional mean estimator for an infinite number of mixture components, and at the same time it was shown that a small number of components achieves a good performance in practice. The idea is to fit a GMM to channel data, which is used afterwards for channel estimation based on noisy observations. The membership to a mixing component in a GMM can be interpreted as a discrete hidden or latent variable. 
In contrast, the variational autoencoder (VAE) method~\cite{Kingma2014,Rezende2014}, which shares common features with the GMM approach, uses a continuous latent space to learn the distribution of the underlying data. In a variety of tasks in communications, the VAE showed its effectiveness, for instance in channel equalization~\cite{Caciularu2018}. However, to our knowledge, there are no approaches in the literature that use the VAE for channel~estimation.

Our contributions in this work are as follows. We propose a framework that allows to employ the VAE for channel estimation. To this end, we train a VAE to generate a conditional mean and covariance for each channel realization, which can subsequently be used to determine the individual linear minimum mean square error (LMMSE) channel estimator for a given observation of the channel.
Due to the inherent structure of the VAE, the probability distribution of the underlying radio scenario is approximated as a conditionally Gaussian distribution.
The VAE thus provides an overall prior information about channel state conditions. 
Further analysis shows that the VAE channel estimator is able to asymptotically approximate the MMSE estimator.
In total, we propose three variants of VAE-based channel estimators. 
We first analyze the full potential of the VAE by allowing genie-knowledge of the true underlying channel for the latent encoding. We then propose practically feasible approaches that either use noiseless channel data solely in the training phase or only work with noisy data for both training and testing.
Our simulations highlight that these approaches provide strong channel estimators close to the genie-based estimator, which offers a lot of potential, also for a prospective application to other system models.

%% file: content/system.tex
\section{System and Channel Model}
\label{sec:system}

In this work, we consider a single-input multiple-output (SIMO) communications scenario. A base station (BS) equipped with $M$ antennas receives uplink training signals from a single antenna mobile terminal (MT). In particular, at the BS after decorrelating the pilot signal, noisy observations
\begin{equation}
    \vy = \vh + \vn \in \CC^M
    \label{eq:system}
\end{equation}
are received, where the channel $\vh$ is perturbed with Gaussian noise $\vn\sim\mathcal{N}_{\CC}(\bm{0},\msig)$. We assume that the BS is equipped with a uniform linear array (ULA) with half-wavelength spacing. The channel covariance of a ULA is known to have a Toeplitz structure, which can be asymptotically approximated by a circulant matrix for a large number of antennas~\cite{Gr06}. Diagonalization of a circulant matrix $\mc$ can be obtained with the help of the discrete Fourier transform (DFT) matrix $\mf\in\CC^{M\times M}$: 
\begin{equation}\label{eq:circ-decomp}
    \mc=\mf\herm\diag(\vc)\mf, \qquad \vc\in\RR^M.
\end{equation}
In Section~\ref{sec:vae-ce}, we will use this circulant approximation to design a channel estimator in three variations.

Moreover, we consider different channel models in this work to validate our method. The 3rd Generation Partnership Project (3GPP) defines spatial channel models which allow the channel to be modeled as~\cite{3GPP2020}
\begin{equation}
    \vh\mid \vdel \sim \mathcal{N}_{\CC}(\bm{0}, \mc_{\vdel})
    \label{eq:cond-gauss}
\end{equation}
where the random vector $\vdel$ encodes the information about path gains and angles of arrival that define the propagation clusters that belong to the channel. Since we consider a SIMO scenario here, the covariance 
\begin{equation*}
    \mc_{\vdel}=\int_{-\pi}^{\pi} g(\vartheta;\vdel) \va(\vartheta)\va(\vartheta)\herm \diff \vartheta
\end{equation*}
represents the receive-side covariance. The terms $g(\vartheta;\vdel)\geq 0$ and $\va(\vartheta)$ depict a power angular spectrum and the array steering vector, respectively. It has to be noted that, although the channel is modeled as conditionally Gaussian, without knowledge of $\vdel$, which cannot be assumed in practice, the channels are clearly not Gaussian distributed. Also, we consider a single snapshot scenario, where for each channel realization an independent realization of $\vdel$ is drawn.

More realistic channels can be obtained with the QuaDRiGa channel simulator~\cite{Jaeckel2014,QUADRIGA2016}. Therein, channels are modeled as a superposition of $L$ propagation paths such that 
\begin{equation*}
    \vh = \sum_{\ell=1}^L \vg_\ell \exp(-2\pi \jim f_c \tau_\ell)
\end{equation*} 
where the carrier frequency is denoted as $f_c$ and the delay of the $\ell$-th path as $\tau_\ell$. The vector $\vg_\ell$ expresses the attenuation characteristics between every antenna pair, as well as the antenna radiation pattern and polarization. After generation, the channels are post-processed to remove the path gain, as described in the QuaDRiGa documentation~\cite{QUADRIGA2016}. 

For both 3GPP and QuaDRiGa specifications, we create a training dataset consisting of $N_t=10^5$ samples and a test dataset consisting of $N_v=10^4$ samples.

%% file: content/model.tex
\section{VAE for Channel Estimation}
\label{sec:vae-ce}

The aim of this section is to briefly introduce the concept behind the VAE and which objective it follows. Afterwards, we present an approach how a VAE, that is trained on channel data in a completely unsupervised manner, can be leveraged for channel estimation.


\subsection{VAE Preliminaries}
\label{subsec:vae}

Variational Inference (VI) builds the foundation of the VAE where the optimization of the evidence lower bound (ELBO) is the central objective~\cite{Blei2017}. A well-known decomposition of the log-likelihood of a data point $\vx\sim p(\vx)$ reads as
\begin{equation}
	\log p(\vx) = \mathcal{L}(q) + \KL(q(\vz|\vx)\,\|\,p(\vz|\vx))
	\label{eq:log-like}
\end{equation}
with the ELBO $\mathcal{L}(q)$ and the Kullback-Leibler divergence 
\begin{equation*}
	\KL(q(\vz|\vx)\,\|\,p(\vz|\vx)) = \E_{q(\vz|\vx)}\left[\log\frac{q(\vz|\vx)}{p(\vz|\vx)}\right].
\end{equation*}
VI introduces the variational distribution $q(\vz|\vx)\in\mathcal{Q}$, which belongs to a family of distributions $\mathcal{Q}$, and depends on the so-called latent vector~$\vz$. Its purpose is to approximate the true posterior distribution $p(\vz|\vx)$ as closely a possible. A maximization of the ELBO achieves two goals: \textit{1)} the data log-likelihood is maximized, and \textit{2)} the KL divergence between $q(\vz|\vx)$ and the posterior $p(\vz|\vx)$ is minimized.

\begin{figure}[t]
	\begin{minipage}[b]{1.0\linewidth}
		\centering
		\centerline{\includegraphics{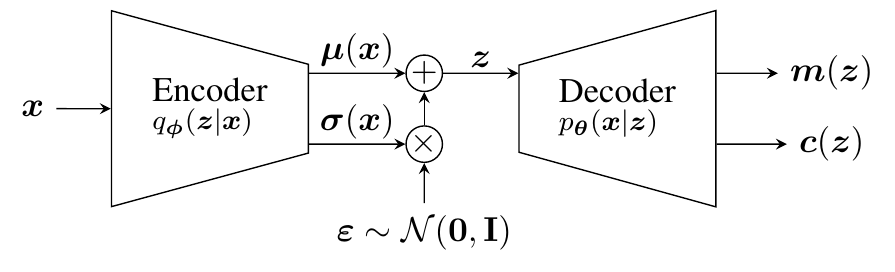}}
	\end{minipage}
	\vspace{-7mm}
	\caption{Structure of a VAE with Gaussian posteriors for $q_{\phi}(\vz|\vx)$ and $p_{\theta}(\vx|\vz)$. The encoder and decoder represent neural networks.}
	\label{fig:vae}
	\vspace{-3mm}
\end{figure}

The VAE optimizes the ELBO with the help of neural networks and the reparameterization trick~\cite{Kingma2014,Rezende2014}. This makes VI broadly accessible as it enables to process large amounts of data in a straight-forward way. The drawing in Fig.~\ref{fig:vae} illustrates the functionality of the VAE. For training of a VAE, it is beneficial to express the ELBO as
\begin{equation}
	\mathcal{L}(q) = \E_{q_{\phi}(\vz|\vx)}\left[\log p_{\theta}(\vx|\vz)\right] - \KL(q_{\phi}(\vz|\vx)\,\|\,p(\vz)).
	\label{eq:vae}
\end{equation}
The parameters $\theta$ and $\phi$ highlight certain parameterizations for the posteriors $p_{\theta}(\vx|\vz)$ and $q_{\phi}(\vz|\vx)$, respectively. Their technical implementation is represented by the encoder and decoder of the VAE.

At this point, we must decide for a type and structure of the distributions that appear in~\eqref{eq:vae}. A common choice for this is to assume diagonal Gaussian distributions.
Consequently, the probability distributions involved are 
\begin{IEEEeqnarray*}{lll}
	& p(\vz) & =\mathcal{N}(\bm{0},\mathbf{I}), \\
	& p_{\theta}(\vx|\vz)\ & =\mathcal{N}_{\CC}(\vm(\vz),\diag(\vc(\vz))), \\  
	\text{and}\ & q_{\phi}(\vz|\vx) & =\mathcal{N}(\vmu(\vx),\diag(\vsig^2(\vx))).\label{eq:z_given_x}
\end{IEEEeqnarray*}
With these definitions, let us examine Fig.~\ref{fig:vae} in more detail. The encoder takes a data sample $\vx$ and maps it to the mean value $\vmu(\vx)$ and standard deviation vector $\vsig(\vx)$ of $q_{\phi}(\vz|\vx)$. A sample $\veps$ from a standard Gaussian distribution in combination with $\vmu(\vx)$ and $\vsig(\vx)$ is used to obtain a reparameterized sample $\vz=\vmu(\vx)+\vsig(\vx)\times\veps$, the latent variable. The sample $\vz\in\RR^L$ is fed into the decoder in the last step, to obtain the mean value $\vm(\vz)$ and covariance vector $\vc(\vz)$ of the Gaussian distribution $p_{\theta}(\vx|\vz)$.

Additionally, the usage of diagonal Gaussians lets us find closed-form and simple-to-compute expressions for the terms in the optimization objective in~\eqref{eq:vae}. The expectation on the left can be approximated with Monte-Carlo samples $\vz^{(k)}\sim q_{\phi}(\vz|\vx), k=1,\ldots,K,$ such that $\E_{q_{\phi}(\vz|\vx)}\left[\log p_{\theta}(\vx|\vz)\right]$ is
\begin{equation}
	\sum_{k=1}^K \left\{ -M\log\pi - \sum_{m=1}^M \log c_i(\vz^{(k)}) - \frac{|x_i-m_i(\vz^{(k)})|^2}{c_i(\vz^{(k)})} \right\}.
	\label{eq:vae-dec-like}
\end{equation}
The subscript $i$ symbolizes that the $i$-th element of the respective vector is selected, e.g., $x_i$ is the $i$-th element of $\vx$. The KL-term  $\KL(q_{\phi}(\vz|\vx)\,\|\,p(\vz))$ results in
\begin{equation}
	\frac{1}{2} \left( \sum_{l=1}^{L} \left\{ -\log\sigma^2_i(\vx) + \mu_i(\vx)^2 + \sigma^2_i(\vx) \right\} - L \right).
	\label{eq:vae-kl}
\end{equation}

It should be noted that we can approximate the true covariance of the channels with the presented VAE. This is the case because the covariance is Toeplitz structured, which can be approximated by a circulant matrix as described in Section~\ref{sec:system} and~\cite{Gr06,Baur2022a}. Suppose now that the VAE receives the input $\vx=\vh$. 
We would like to have the corresponding channel covariance of the form~\eqref{eq:circ-decomp} at the decoder output. This, however, would require unconstrained (and in particular non-diagonal) covariance matrices at the decoder output. As a remedy, we Fourier-transform all channel samples and feed \( \vx = \mf \vh \) into the encoder because the corresponding channel covariance is then given be the diagonal matrix \( \diag(\vc) \) which can be readily modeled by the decoder.


\subsection{MMSE Approximation of the VAE Channel Estimator}
\label{subsec:conjecture}
	In this section, we want to motivate to use the VAE for channel estimation. To this end, we rewrite the MSE-optimal conditional mean estimator and find structural relations to the VAE, resulting in an approximation of the optimal channel estimator as we discuss in the remainder of this section.
	
	For any arbitrarily distributed random variable \( \vh \), we can always find a condition which makes it Gaussian.
	For example, if \( \bm{t} \) is a Gaussian random variable, we can write \( \vh = (\vh - \bm{t}) + \bm{t} \) and \( \vh \mid (\vh - \bm{t}) \) is Gaussian.
	Finding a suitable condition can be challenging---in particular if the distribution of \( \vh \) is not known.
	In the example, the condition depends on \( \vh \) itself.
	The VAE's goal is to achieve a conditional Gaussianity via the latent variable \( \bm{z} \) so that we have
	\begin{equation}\label{eq:cond_gauss_vae}
		\vh \mid \vz \sim \mathcal{N}_{\CC}(\vmu_{\vh\mid \vz}, \mc_{\vh\mid\vz}).
	\end{equation}
	It is known that MMSE channel estimates are given by the conditional mean $\E[\vh\mid \vy]$. The law of total expectation lets us write
	\begin{equation}\label{eq:total_exp}
		\E[\vh\mid\vy] = \E_{\vz}[\E[\vh\mid\vz,\vy]\mid\vy].
	\end{equation}
	Assuming the VAE achieves its goal and~\eqref{eq:cond_gauss_vae} holds,
	we have a closed-form expression for the inner expectation due to the conditional Gaussianity~\cite{Kay1993}:
	\begin{equation}\label{eq:lmmse}
		\E[\vh\mid \vz,\vy] = \vmu_{\vh\mid \vz} + \mc_{\vh\mid \vz} (\mc_{\vh\mid \vz} + \msig)^{-1} (\vy - \vmu_{\vh\mid \vz}).
	\end{equation}
	Note that $\vz$ depends on $\vy$ through the encoder and $\vmu_{\vh\mid \vz}$ and $\mc_{\vh\mid \vz}$ are evaluated by the decoder of the VAE. The outer expectation in \eqref{eq:total_exp} depends on the conditional distribution of $p(\vz|\vy)$ that is approximated by $q_\phi(\vz|\vx)$, which is used to evaluate~\eqref{eq:total_exp} in our method. The vector $\vx$ is the input of the encoder, whose design is discussed in detail in Section~\ref{subsec:vae-ce}, but it should be noted that $\vx$ is a transformed version of $\vy$. For simplicity, assume $\vx=\vy$ until further notice for the remainder of Section~\ref{subsec:conjecture}. It remains to compute the outer expectation in~\eqref{eq:total_exp} after plugging in~\eqref{eq:lmmse}.
	
	To this end, recall the encoder's objective (see also Fig.~\ref{fig:vae}):
	If \( \bm{x} \) is its input,
	the encoder computes a mean \( \bm{\mu}(\bm{x}) \) and a variance \( \bm{\sigma}^2(\bm{x}) \) such that the latent representation \( \vz = \bm{\mu}(\bm{x}) + \bm{\varepsilon} \times \bm{\sigma}(\bm{x}) \) is obtained, which embodies samples from $q_\phi(\vz|\vx)$.
	If we furthermore assume that $p(\vz|\vy)$ is perfectly represented by $q_\phi(\vz|\vx)$, we can approximate the outer expectation in~\eqref{eq:total_exp} using samples of the form \( \vz^{(k)} = \bm{\mu}(\bm{x}) + \bm{\varepsilon}^{(k)} \times \bm{\sigma}(\bm{x}) \) where every \( \bm{\varepsilon}^{(k)} \) is a sample of \( \bm{\varepsilon} \sim \mathcal{N}(\bm{0},\mathbf{I}) \).
	With sufficiently many samples $\vz^{(k)}$, $k=1,\dots,K$, we are able to asymptotically approximate the MMSE channel estimator as a direct consequence of the law of large numbers~\cite{Loeve1977}, i.e.,
	\begin{equation}
		\E[\vh\mid\vy] \approx \frac{1}{K} \sum_{k=1}^K \E[\vh\mid \vz^{(k)},\vy].
		\label{eq:conjecture}
	\end{equation}
	Note that the channel estimates on the right-hand side can still be efficiently computed with the VAE in combination with~\eqref{eq:lmmse}. To save computational complexity, we use only one sample---namely \( \vmu(\vx) \)---in our numerical experiments and still achieve remarkable results. Under this assumption, the estimator from \eqref{eq:conjecture} simplifies for $K=1$ as
	\begin{equation}
		\E[\vh\mid\vy] \approx \E[\vh\mid \vz^{(1)}=\vmu(\vx),\vy].
		\label{eq:lmmse-mu}
	\end{equation}
	
	For another motivation for approximating the outer expectation in~\eqref{eq:total_exp} using only the mean \( \bm{\mu}(\bm{x}) \),
	we can interpret \( \bm{\sigma}(\bm{x}) \) as the uncertainty of the encoder to produce a suitable latent representation.
	If the noise variance in~\eqref{eq:system} is small, we expect this uncertainty to be small so that all samples are close to \( \bm{\mu}(\bm{x}) \) and \( \bm{z} \approx \bm{\mu}(\bm{x}) \) holds.


	\subsection{Channel Estimation with the VAE}
	\label{subsec:vae-ce}
	A common assumption, which we make here as well, is white noise, which yields $\msig=\varsigma^2\mathbf{I}$, with the noise variance $\varsigma^2$ that is given in our experiments. For the LMMSE estimator in~\eqref{eq:lmmse}, it is necessary to know the conditional mean $\vmu_{\vh\mid \vz}$ and covariance $\mc_{\vh\mid\vz}$ of $\vh$ given $\vz$ to estimate the channel based on $\vy$. A close look at Fig.~\ref{fig:vae} reveals how the VAE can be used for channel estimation. For an input sample $\vx$, the VAE decoder delivers a conditional mean $\vm(\vz)$ and a diagonal covariance $\vc(\vz)$ that parameterize the distribution in \eqref{eq:cond_gauss_vae}. 
	As we use Fourier-transformed input samples for complexity savings as explained before, the quantities in~\eqref{eq:lmmse} of the conditionally Gaussian are computed as
	\begin{equation}
		\vmu_{\vh\mid \vz} = \mf\herm \vm(\vz), \quad \mc_{\vh\mid\vz} = \mf\herm\diag(\vc(\vz))\mf.
		\label{eq:cond_gauss_mean_cov}
	\end{equation}
	In the following, we present three possible channel estimators that leverage the VAE. All three estimators have in common that the VAEs can be trained offline before their application to channel estimation. Moreover, all estimators use $\vz=\vmu(\vx)$ as in~\eqref{eq:lmmse-mu} during the estimation phase after training.
	
	\textit{1):} The transformed true channel $\vx=\mf\vh$ is fed into the VAE encoder for both training and evaluation. Its latent representation is the basis for the evaluation of the decoder likelihood model $p_{\theta}(\vx|\vz)$ in~\eqref{eq:vae-dec-like}. The outputs of the encoder, i.e., $\vmu(\vx)$ and $\vsig(\vx)$, are used to evaluate the KL-term in~\eqref{eq:vae-kl} during training.
	This estimator is supposed to produce the best estimation results because conditional mean and covariance at the decoder are inferred with the true channel at the encoder and its latent representation. The relation $\vz=\vmu(\vx)$ holds well in this situation as the noise variance is zero. In our simulations, we could also observe that the variance $\vsig^2(\vx)$ at the encoder is approximately zero which is a strong motivation for the usage of the estimator in~\eqref{eq:lmmse-mu}. Although this estimator even has the potential to outperform the conditional mean estimator $\E[\vh\,|\,\vy]$, as the true channel state information acts as side information, this estimator is obviously not applicable to a real scenario since it requires knowledge of the channel during the evaluation phase. Instead, it inspires the basic idea of the proposed method and can serve as a suitable benchmark result in a channel estimation scenario where the optimal estimator itself is unknown and inaccessible. We therefore call this estimator \textit{VAE-genie}.
	
	\textit{2):} The decoder likelihood $p_{\theta}(\vx|\vz)$ in~\eqref{eq:vae-dec-like} is again computed with the true $\vx = \mf\vh$ during training, but the encoder is fed with $\tilde{\vx}=\mf\vy$. This causes the encoder to output $\vmu(\tilde{\vx})$ and $\vsig(\tilde{\vx})$ based on which the KL-term in~\eqref{eq:vae-kl} is evaluated during training. The adaption also changes the variational distribution from $q_{\phi}(\vz|\vx)$ to $q_{\phi}(\vz|\tilde{\vx})$. Practical deployment of this estimator is now possible as the encoder's input is $\tilde{\vx}$, which allows to use this VAE for estimation of $\vh$ based on $\vy$ after training.
	The relation $\vz=\vmu(\vx)$ holds only for high SNR. At low SNR, we therefore expect a performance loss when using only $\vmu(\vx)$ in~\eqref{eq:lmmse-mu}. Moreover, it is expected that this estimator delivers worse estimation quality than VAE-genie as is does not have access to the true channel in the evaluation phase. It is, in contrast, applicable to a real scenario, as the true channels are only used during training. We call this estimator \textit{VAE-noisy}.
	
	\textit{3):} The VAE is fed with $\tilde{\vx}=\mf\vy$ at the encoder and~\eqref{eq:vae-dec-like} is also evaluated with $\tilde{\vx}$ during training. The KL-term in~\eqref{eq:vae-kl} is again evaluated with the decoder outputs $\vmu(\tilde{\vx})$ and $\vsig(\tilde{\vx})$ during training. The variational distribution stays the same as for VAE-noisy, but the decoder likelihood changes from $p_{\theta}(\vx|\vz)$ to $p_{\theta}(\tilde{\vx}|\vz)$. It follows that this model learns a lower bound to $p(\tilde{\vx})$, which is different from our actual objective. For $\vmu_{\vh\mid \vz}$ in~\eqref{eq:lmmse}, this is not a problem as long as $\E[\vn]=\bm{0}$, which is the case in~\eqref{eq:system}. The channel covariance can be determined with a simple workaround. While the VAE decoder continues to output $\vc(\vz)$, the term $\diag(\vc(\vz))+\varsigma^2\mathbf{I}$ is used instead to compute~\eqref{eq:vae-dec-like} during training. In this way the decoder's learning algorithm is forced to substitute only the desired part, the diagonal channel covariance $\vc(\vz)$. The mean is left as $\vm(\vz)$. 
	Note that this adaptation is also necessary for the back-transformation in \eqref{eq:cond_gauss_mean_cov}.
	In this way, we enforce that the decoder outputs the channel covariance, by training solely on noisy observations $\vy$ at the encoder and decoder. As for VAE-noisy, the relation $\vz=\vmu(\vx)$ holds only for high SNR and we expect to observe a similar performance loss at low SNR by only using $\vmu(\vx)$ in~\eqref{eq:lmmse-mu}. It should be noted that no access to the true channels is needed, neither during training nor during evaluation. We call this more practical and realistic estimator \textit{VAE-real}.
	
	Finally, we want to state that evaluating~\eqref{eq:lmmse-mu} with more samples than exclusively with $\vmu(\vx)$, would certainly deliver better estimation results, especially for VAE-noisy and VAE-real at low SNR. It is of great interest to know how many samples are necessary for best estimation results. Our future work should therefore also cover such an analysis.


	\subsection{VAE Architecture}
	\label{subsec:architecture}
	
	Since the channels are complex-valued, we stack the real and imaginary part of the vector channels to create input vectors of size $2M$ for the neural networks of the VAE. Please note, the test dataset is previously unseen data for the VAE. The neural networks used in every VAE have the following architecture: encoder and decoder are built up symmetrically. The encoder consists of three times a building block of a convolutional (conv.) layer, Batch Normalization (BN) layer, and ReLU activation function. 
	Each conv. layer uses a kernel of size 7, the input samples are mapped from 1 to 8, to 32, to 128 conv. channels, and a stride of 2 is used. 
	The three blocks are followed by two linear layers that map to $\vmu(\vx)$ and $\vsig(\vx)$. The decoder architecture is analog to the encoder architecture, just flipped symmetrically. We use a learning rate of $10^{-4}$ during the training phase. The architectures were found with a random search over the hyperparameter space by searching for the configuration that yields the highest ELBO value. 
	Our neural networks are implemented with the help of \textit{PyTorch}, and optimized with the \textit{Adam} optimizer~\cite{Kingma2015}, together with the method of free bits~\cite{Kingma2019}. 
	Note that during training of VAE-noisy and VAE-real we create new noisy observations $\vy$ in every epoch by sampling the additive noise term in \eqref{eq:system}. This exhibited an additional performance gain in our experiments. The VAEs are trained exclusively for a certain SNR value, except VAE-genie where this is not required. We train our VAEs until the ELBO saturates and use the model that yields the highest ELBO value. 
	

	\subsection{Related Channel Estimators}
	\label{subsec:otherce}
	
	For performance evaluation of our proposed estimators, we compare them with other channel estimators from the literature. In the case of 3GPP channel data, we have access to the covariance matrix $\mc_{\vdel}$ of the conditionally Gaussian channel in~\eqref{eq:cond-gauss}. This lets us evaluate the conditional mean estimator~\cite{Neumann2018}, which is given by
	\begin{equation}
		\hat{\vh}_{\text{genie-cov}} = \mc_{\vdel} (\mc_{\vdel} + \msig)^{-1} \vy,
		\label{eq:genie-cov}
	\end{equation}
	and represents the true MMSE estimator for the conditionally Gaussian channel.
	Genie-knowledge in the form of $\mc_{\vdel}$ is required for this estimator, which is why we call it \textit{genie-cov}. Note that it cannnot be used for QuaDRiGa based channel data. Another closely related estimator is the evaluation of~\eqref{eq:genie-cov} with a sample covariance matrix. To this end, a covariance matrix 
	\begin{equation*}
		\hat{\mc}=\frac{1}{N_v}\sum_{i=1}^{N_v} \vh_i\vh_i\herm
	\end{equation*}
	is computed from the $N_v$ samples $\vh_i$ in the evaluation dataset, and $\hat{\mc}$ is used in~\eqref{eq:genie-cov} instead of $\mc_{\vdel}$. We refer to this estimator as \textit{sample-cov}. A simple, yet commonly employed estimator, is based on least squares (LS), i.e., $\hat{\vh}_{\text{LS}}=\vy$. 
	We also compare our approaches to the GMM-based channel estimator~\cite{Koller2021icassp, Koller2022}. We expect that the GMM estimator is the strongest baseline we compare our VAE channel estimation approaches to as it is proven to be asymptotically optimal for an infinite number of mixture components. The GMM is fitted with 128 components and the covariance matrix of each component is set to be circulant, so that both methods produce circulant covariances.

%% file: content/results.tex
\section{Simulation Results}
\label{sec:results}

This section presents the channel estimation results of our numerical simulations based on 3GPP and QuaDRiGa channel data. In all our experiments, we calculate the normalized mean squared error (NMSE) as 
\begin{equation*}
	\text{NMSE} = \frac{1}{N_v}\sum_{i=1}^{N_v}\frac{\|\vh_i-\hat{\vh}_i\|^2}{\|\vh_i\|^2}
\end{equation*}
for the evaluation dataset. We define the SNR on a per-sample basis such that $\text{SNR}=\|\vh\|^2/(M\varsigma^2)$. 

\begin{figure}[!t]
	\hspace{-10pt}
	\includegraphics{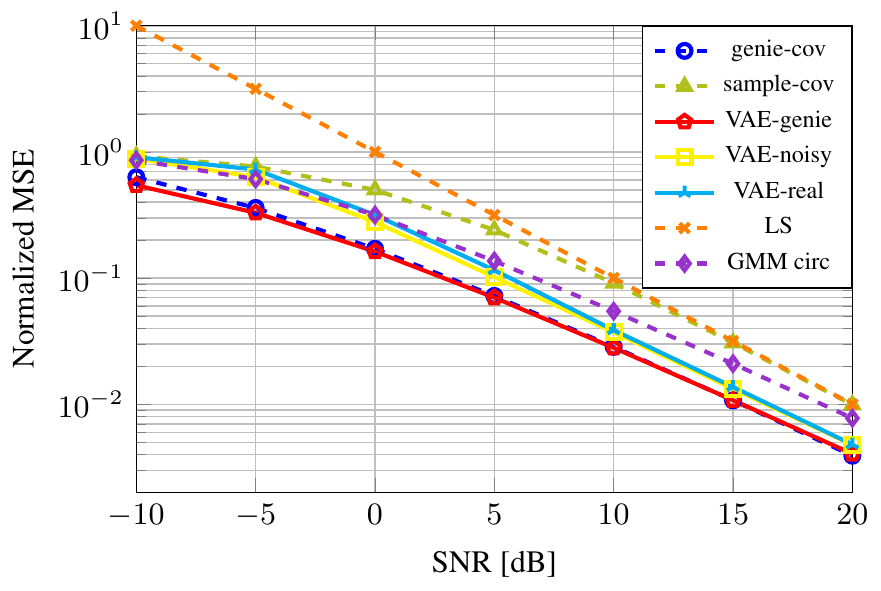}
	\vspace{-3mm}
	\caption{NMSE over SNR for 3GPP channel data with 3 propagation clusters and 32 antennas at the BS. Our approaches are displayed as solid curves.}
	\label{fig:3gpp_32}
	\vspace{-2mm}
\end{figure}

\begin{figure}[!t]
	\hspace{-10pt}
	\includegraphics{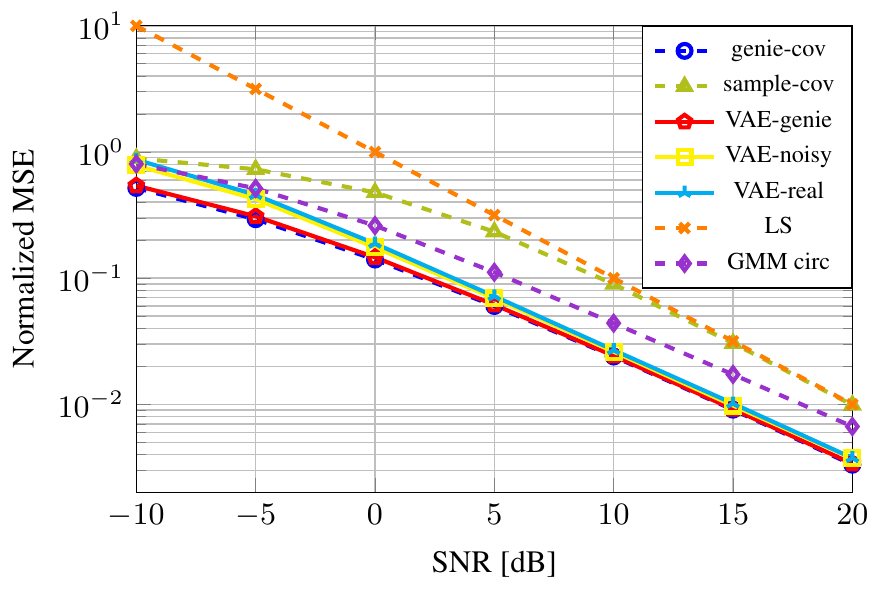}
	\vspace{-3mm}
	\caption{NMSE over SNR for 3GPP channel data with 3 propagation clusters and 128 antennas at the BS. Our approaches are displayed as solid curves.}
	\label{fig:3gpp_128}
	\vspace{-2mm}
\end{figure}

\begin{figure}[!t]
	\hspace{-10pt}
	\includegraphics{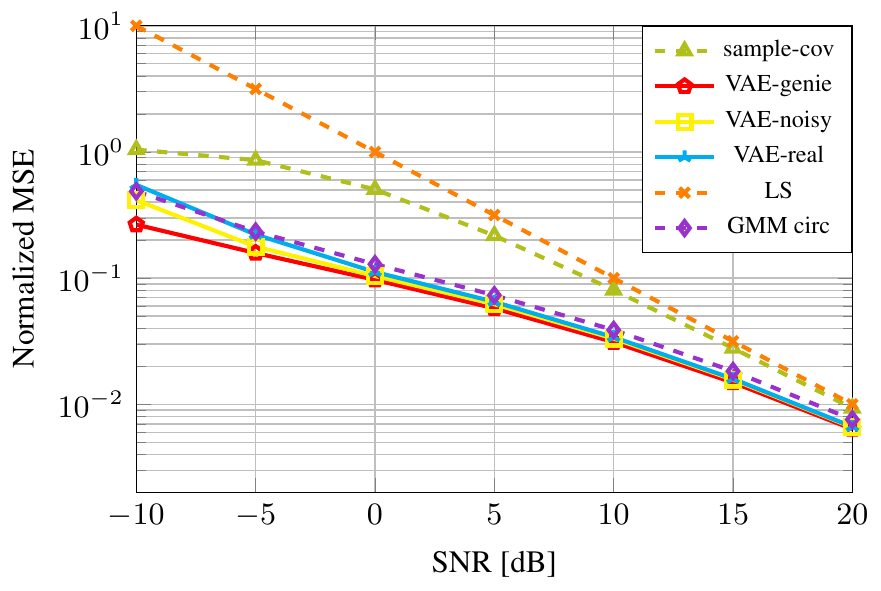}
	\vspace{-3mm}
	\caption{NMSE over SNR for QuaDRiGa channel data UMa only LOS with 128 antennas at the BS. Our approaches are displayed as solid curves.}
	\label{fig:quadriga_128}
	\vspace{-2mm}
\end{figure}

Fig.~\ref{fig:3gpp_32} shows the NMSE over a SNR from -10 to 20\,dB for 3GPP data with 3 propagation clusters and 32 antennas at the BS. Path gains are sampled from a uniform distribution in $[0,1]$ to sum up to one with a maximum difference of 9\,dB. Angles of arrival are also sampled from a uniform distribution in $[\ang{-90},\ang{90}]$ with a minimum difference of $\ang{1}$. In this plot and the following, we display our approaches with solid, all others with dashed linestyles. The plot shows that VAE-genie is almost everywhere on par with genie-cov. In the low SNR regime it is even possible to beat genie-cov. 
The explanation for this at first glance unintuitive behavior has already been mentioned before when VAE-genie was introduced. 
This method uses the information about the actual channel state as side information to achieve an optimal representation of the channel in the latent space of the VAE, and thus has even more information than the conditional mean estimator.
The worst performance in Fig.~\ref{fig:3gpp_32} shows LS, followed by sample-cov. This is expectable as these approaches are not scenario-specific and very simple. VAE-noisy is always better than VAE-real, which is a reasonable behavior since VAE-noisy has access to the noiseless channels, but the performance gap is small. At high SNR, VAE-real converges to VAE-noisy as $\vy$ converges to $\vh$. At low SNR, VAE-noisy has a few dB advantage over VAE-real. The GMM is better than the non-genie VAEs at low SNR, and becomes worse from 0\,dB on. All displayed methods, except genie-cov, stay clearly behind the performance of VAE-genie.

The results for 3GPP data with 3 propagation clusters and 128 antennas at the BS in Fig.~\ref{fig:3gpp_128} are in agreement with the findings for 32 antennas at the BS. Simulation settings are identical to the 32 antennas case. Fig.~\ref{fig:3gpp_128} shows that VAE-genie is not able to beat genie-cov in this setting, but almost for every SNR value it is on par with genie-cov. 
The sample-cov and LS approach still perform worst. VAE-noisy and VAE-real perform similarly as in Fig.~\ref{fig:3gpp_32}, but manage to reach the genie curves more closely for higher SNR values. This can be explained with the asymptotic equivalence of the Toeplitz and circulant matrix for higher dimensions \cite{Gr06}. The GMM shows a similar performance as in Fig.~\ref{fig:3gpp_32}. In consequence of the better performance for higher SNR of VAE-noisy and VAE-real, the performance of the GMM appears worse than in the 32 antenna setting when compared to the mentioned VAE methods.

At this point, we want to clarify that the conditional Gaussianity of the 3GPP channels in \eqref{eq:cond-gauss} is not connected to the objective of the VAE to model the channels conditionally Gaussian as stated in \eqref{eq:cond_gauss_vae}. More precisely, the latent variable $\vz$ of the VAE is not trained to approximate $\vdel$ in \eqref{eq:cond-gauss}. In fact, the VAE channel estimator is not based on a specific channel model since any random variable can be modeled as conditionally Gaussian as reasoned in \ref{subsec:conjecture}.
To highlight this, we investigate the channel estimation quality of our methods based on QuaDRiGa channel data. 
In contrast to the 3GPP model from before, the channels are not explicitly modeled as conditional Gaussians.
The results for a 3GPP 38.901 urban macrocell (UMa) scenario with only LOS channels and single-antenna users (80\,\% of them indoors), cf. \cite{QUADRIGA2016}, with 128 antennas at the BS are displayed in Fig.~\ref{fig:quadriga_128}.
Note that no true covariance is available in this case, which is why no genie-cov curve is plotted. 
The behavior of the curves is similar to the behavior observable in the 3GPP plot with 128 antennas. VAE-genie always shows the best performance, although VAE-noisy and VAE-real approach it very closely from 0\,dB on. The GMM is closer to the performance of the VAEs in this case. The sample-cov and LS approach are still the worst performing methods with a large gap to the proposed estimators. 
This QuaDRiGa experiment highlights that our VAE based channel estimation methods do not only perform well when we have conditionally Gaussian modeled channels, but also for general channel models. Moreover, in cases where the optimal estimator is not available, the proposed VAE-genie delivers a data-based approach that can serve as a benchmark which might be helpful for evaluating various applications.

%% file: content/conclusion.tex
\section{Conclusion}
\label{sec:conclusion}

In this work, we presented a novel approach for data-driven channel estimation. Its basis forms a VAE that enables us to learn a conditional mean and covariance for every channel that can subsequently be used in the conditional mean channel estimation formula. 
Due to the inherent structure of the VAE it approximates the MMSE estimator under certain assumptions, regardless of the channel distribution. Simulation results demonstrated this supposed optimality for different channel models and highlighted the potential of our methods. 
Two of them deserve a special mentioning. The VAE-genie may be used as a baseline for channel estimation in a communications scenario, while the VAE-real can be trained solely with noisy observations without access to a dataset of noiseless channels. We find that the VAE-real is a particularly interesting method because noiseless channel data are, in reality, unavailable, as there always remains estimation noise in the data. Our future work includes a rigorous proof of the conjectured MMSE approximation of the proposed techniques, an investigation of how many samples are necessary in~\eqref{eq:conjecture}, and the extension of our methods to other system models. 